\documentstyle[preprint,aps]{revtex}
\begin {document}
\draft
\preprint{UCI TR 93-32}
\title{On the Distribution of Neutral and Charged Pions\\through the
Production of a Classical Pion Field}
\setcounter{footnote}{0}
\author{A.\ A.\ Anselm \footnote{Electronic address:
anselm@lnpi.spb.su}
}
\address{
Petersburg Nuclear Physics Institute, Gatchina, 188 350 St.
Petersburg, Russia}
\author{Myron Bander\footnote{Electronic address:
mbander@funth.ps.uci.edu}
}
\address{
Department of Physics, University of California, Irvine, California
92717, USA}

\maketitle
\begin{abstract}
High energy reactions may produce a state around the collision point that is
best described by a classical pion field. Such a field might be an isospin
rotated vacuum of the chiral $\sigma$-model or, as discussed in this work, a
solution of the equations of motion resultinng from the coupling of
fields of this model to quarks produced in the collision. In such
configurations all directions in isospin space are allowed leading to a
sizable probability  of events with, essentially, only charged particles
(Centauros) or all neutral particles (anti-Centauros). (In more common
statistical models of multiparticle production, the probability of such
events  is suppressed exponentially by the total multiplicity.) We find that
the isospin violation due to the mass difference of the up and down quarks
has a significant effect on these distributions and enhances the production
of events consisting  predominantly of neutral particles.
\end{abstract}
\narrowtext
{\bf 1.} In recent years several authors
\cite{Anselm,AnsRys,BlaKrz,bj1,bj2,RajWil}
have suggested that the celebrated Centauro events \cite{Lattes}, in
which no $\pi^0$'s have been observed versus a large number of charged
hadrons, might be explained by the production at these high energies
of a classical pion field; an interesting example is the ``disoriented
chiral condensate'' \cite{bj1,bj2}. The idea is that such a process,
considered event-by-event, would correspond to the field being along a
given Cartesian isospin direction. In events where the isospin is
oriented (almost) parallel to the 3-rd axis one would expect mainly
neutral pions while in events where the isospin lies in the
perpendicular plane predominantly charged pions would be produced. Let
$(\pi_1,\pi_2,\pi_3)$ be the three Cartesian isotopic amplitudes of the
classical pion field. As all the orientations are equivalent, the
distribution in the amplitude $\pi_3$ is
\begin{equation}
dw\sim  d\pi_3;\ \ \  \mbox{{\boldmath $\pi$}}^2=\pi_1^2+\pi_2^2
+\pi_3^2=\mbox{const}\, .
\label{cartpi}
\end{equation}
The number of neutral pions, $n_0$, is proportional to $\pi_3^2$
while the total number of produced pions, $n=n_o+n_++n_-\sim
\mbox{{\boldmath $\pi$}}^2$. With $f=n_0/n$, the fraction of neutral
pions, one has from~(\ref{cartpi}),
\begin{equation}
dw=\frac{df}{2\sqrt f}\, ;\label{fdist}
\end{equation}
this distribution is normalized to unity.

Obviously (\ref{fdist}) predicts many more events with a small number
of neutrals than do usual statistical mechanisms for pion production.
In the latter case one expects $dw/df$ to peak at $f=1/3$
($n_+=n_-=n_0=1/3$ as $n\rightarrow\infty$)) and to decrease
exponentially with $n$ as $f$ deviates from this value. The
distribution (\ref{fdist}) corresponds to the limit
$n\rightarrow\infty$ and gives for the relative number of events with
the fraction of neutrals less than $f$
\begin{equation}
P(f)=\int_0^f \frac{dw}{df'}\  df'={\sqrt f}\, .\label{Pof-f}
\end{equation}
For a typical Centauro event $f\sim 1/100$ and $P\sim 10\% $. This seems
to be a reasonable number as the five ``classic'' Centauros represent
about $1\%$ of events with appropriate energies \cite{Lattes}.

At the other end of the spectrum, near $f=1$, the probability of an event
having an anomalously large fraction of $\pi^0$'s is
\begin{equation}
1-P(f)=1-\sqrt f\sim \frac{1}{2}(1-f)\, .\label{antiCent}
\end{equation}
We do not have the square root enhancement exhibited in (\ref{Pof-f})
and instead we find a linear dependence at the end of the spectrum;
however, there still is a finite probability of finding events with a large
number of $\pi^0$'s. It is possible that such ``anti-Centauro'' events
have been observed \cite{antiC} and we shall present a mechanism for
enhancing their probability over that of~(\ref{antiCent}).

The distribution (\ref{fdist}) results from exact isospin symmetry. At
the quark level this symmetry is rather strongly violated due to the
up-down quark mass difference, $m_u\ne m_d$. In this Letter we shall
demonstrate that this mass inequality can enhance the probability of
anti-Centauros.

{\bf 2.} A class of solutions for the pion field whose dynamics are
governed by a non-linear chiral Lagrangian was presented in Ref.
\cite{Anselm}. The results of that work may be understood in the
following simple way. The Lagrangian is
\begin{equation}
{\cal L}=\frac{f_\pi^2}{2}\mbox{Tr}\ (\partial_\mu U\partial^\mu U^{\dag})\, ,
\label{Lagr}
\end{equation}
where $f_\pi=93$ MeV and the unitary matrix $U$ is connected to the pion
fields by
\begin{equation}
U=\exp \left (\frac{i\mbox{\boldmath $\tau$}\cdot\mbox{\boldmath
$\pi$}}{f_\pi}\right )\, .
\end{equation}
For the particular form
\begin{equation}
U=\exp [i\tau_3\,  \theta({\bf r},t)] \label{soln}
\end{equation}
the Lagrangian (\ref{Lagr}) leads to the free equation of motion
\begin{equation}
\partial^2\theta=0\, .\label{waveeq}
\end{equation}
For constant unitary matrices $V_L$ and $V_R$ a generalization of
(\ref{soln}) is
\begin{equation}
U=V_L^{\dag}\exp(i\tau_3\, \theta)V_R\, ;\label{solngen}
\end{equation}
this is a general class of solutions which has been studied in Ref.
\cite{Anselm}. All other known solutions \cite{BlaKrz,Enikova} are particular
cases of
(\ref{solngen}).

At large distances from the collision point we require the normal structure
of the vacuum, i.\ e.\ $U=1$. Likewise we will pick solutions in which
$\theta({\bf r},t)\rightarrow 0$ as $r\rightarrow \infty$. This forces
$V_L=V_R$ and the solutions (\ref{solngen}) reduce to isotopic rotations of
(\ref{soln}). In other words, (\ref{solngen}) takes the form
\begin{equation}
U=\exp [i\mbox{\boldmath $\tau$}\cdot\mbox{\boldmath
$n$}\, \theta(x)]\label{disor}
\end{equation}
for some direction $\mbox{\boldmath $n$}$ in isotopic spin space. A
possible scenario for
the production of a classical pion field discussed in \cite{bj1,bj2} is that
inside a certain volume around the collision point a state corresponding to
a constant (in the volume) $\theta$ is produced. This state is degenerate
with the normal vacuum (in the limit $m_\pi=0$) but is rotated with respect
to it in isotopic spin space. In \cite{bj1,bj2} this situation is referred
to as ``disoriented chiral condensate''. It follows from (\ref{disor}) that
any solution of (\ref{waveeq}) describes chiral dynamics.

We now introduce interactions of pions with quarks keeping in mind that the
pion field is the chiral phase of the quark field \cite{Georgi}. In the
presence of pion fields the quark fields should be modified
\begin{eqnarray}
q_L(x)&\rightarrow& \exp [\frac{i}{2}\mbox{\boldmath
$\tau$}\cdot\mbox{\boldmath
$n$}\, \theta(x)] q_L\nonumber\\
&{}&\\
q_R(x)&\rightarrow& \exp [\frac{-i}{2}\mbox{\boldmath
$\tau$}\cdot\mbox{\boldmath
$n$}\, \theta(x)] q_R\, .\nonumber
\end{eqnarray}
The quark mass terms give rise to the quark-pion interaction Hamiltonian
\begin{equation}
{\cal H}=m_u{\bar u}u+m_d{\bar d}d\rightarrow {\bar q}
\exp (\frac{i}{2}\mbox{\boldmath $\tau$}\cdot\mbox{\boldmath
$n$}\, \theta) \left (
m_++m_-\tau_3\right ) \exp (\frac{i}{2}\mbox{\boldmath
$\tau$}\cdot\mbox{\boldmath
$n$}\, \theta )q +\mbox{h. c.}\, ,\label{interaction}
\end{equation}
where $m_\pm =\frac{1}{2}(m_u\pm m_d)$. For the solution (\ref{disor})
\begin{equation}
{\cal H}={\bar q}(m_++m_-\tau_3 )q-(1-\cos\theta){\bar q}\left (m_++m_-n_3
\mbox{\boldmath $\tau$}\cdot\mbox{\boldmath
$n$}\right )q+
\sin\theta {\bar q}i\gamma_5 \left (m_+\mbox{\boldmath
$\tau$}\cdot\mbox{\boldmath $n$}
+m_-n_3\right )q\, .\label{inter2}
\end{equation}

In the normal vacuum (\ref{inter2}) accounts for the pion mass term through
the existence of the chiral condensate $\langle {\bar q}q\rangle\ne
0$. From (\ref{inter2}) one sees that $m_\pi^2=-m_+\langle {\bar q}q\rangle
/f_\pi^2,\ \ \pi=f_\pi\theta$.

The distributions, in the parameters $\theta$ and $\mbox{\boldmath $n$}$, of a
classical pion field produced in
a high energy collision are expected to depend on a production temperature $T$
and have the form
\begin{equation}
dw\sim \int\prod_x d\theta(x) \exp\left (-\frac{1}{T}\int d^3x{\cal H}
\right ) d\mbox{\boldmath $ n$}\delta(\mbox{\boldmath $n$}^2-1)\,
.\label{ndistr}
\end{equation}

If the quark density in the collision is not too high $\langle {\bar
q}q\rangle$ should be set equal to its usual vacuum value. Expanding
around $\theta=0$ (\ref{ndistr}) becomes
\begin{equation}
dw\sim\int\prod_x d\theta(x) \exp\left (-\frac{m_+
|\langle {\bar q}q\rangle|}{2T}\int d^3x\theta^2\right
)d\mbox{\boldmath $ n$}\delta(\mbox{\boldmath $n$}^2-1)\, .
\end{equation}
For $T=T_c\sim 140$ MeV \cite{Bernard} and a volume $V\sim 100$
fm${}^3$ the above is $\exp [-4<\theta^2>]$; large values of $\theta$
will not be excited. However, after the functional $\theta$
integration the distribution in isospin directions remains uniform
leading immediately to (\ref{fdist}).

{\bf 3.} Our critical assumption is that in the high density medium
created by such collisions the quark density and other bilinears in
$q, {\bar q}$ acquire classical values that may be comparable to or
larger than the vacuum chiral condensate $\langle {\bar q}q\rangle
\simeq -(250\ \mbox{MeV})^3$. From the explicit dependence of
(\ref{inter2}) on $n_3$ we see that isospin rotation symmetry is
broken. We consider two possibilities: either
$I(x)=\langle\!\langle{\bar q}\tau_3 q\rangle\!\rangle\ne
0$ or $P(x)=\langle\!\langle{\bar q}i\gamma_5 q\rangle\!\rangle\ne 0$,
in addition to $S(x)=\langle\!\langle{\bar
q}q\rangle\!\rangle\ne 0$ and are sizable. $\langle\!\langle\cdots
\rangle\!\rangle$ denotes the averaging over
quantum fluctuations and we allow for a smooth (on the microscopic
scale) position dependence. The value of $S(x)$ may differ
significantly from the vacuum value of $\langle {\bar q}q\rangle$.

We first consider the first case, $I(x)\ne 0$; although it has less
interesting consequences it is simpler to analyze. The functional integration
over $\theta(x)$ in (\ref{ndistr}) (in the quadratic approximation)
yields
\begin{equation}
dw\sim\frac{1}{\sqrt{|m_+S(x)+m_-I(x)n_3^2|}}dn_3\, .\label{Idist}
\end{equation}
For the dependence of the above on $n_3$ to be significant it is
necessary for the second term in the square root to be comparable in
magnitude to the first one. This is, however, unlikely as their ratio
is (even for $f=n_3^2=1$)
\begin{equation}
\frac{m_-I(x)}{m_+S(x)}=\frac{m_u-m_d}{m_u+m_d}\
\frac{\langle\!\langle u{\bar u}-
d{\bar d}\rangle\!\rangle}{\langle\!\langle u{\bar u}+d{\bar
d}\rangle\!\rangle}\, ;
\end{equation}
with $m_u-m_d/m_u+m_d\sim-0.3$ and the second factor less than unity
the $n_3$ dependence will be insignificant. We reach the same
conclusion if we allow other components of ${\bar q}\mbox{\boldmath
$\tau$}q$ to acquire some classical value.

The situation is significantly different if we assume that $P(x)$ has
a sizable value. Below, we shall return to see whether this is
feasible, but first
discuss the consequences of this assumption. We are now asked to evaluate
\begin{equation}
dw\sim \int\prod_x d\theta(x) \exp\left \{\frac{1}{T}\int d^3x\exp
\left [m_+S(x)(1-\cos\theta)-m_-n_3P(x)\sin\theta\right ]
\right \} d\mbox{\boldmath $n$}\delta(\mbox{\boldmath $n$}^2-1)\,
.\label{Pdistr}
\end{equation}
The exponent has a minimum for a non-zero $\theta$ obtained from
$\tan\theta=m_-n_3P(x)/m_+S(x)$. The functional
integral can be done
(again in a quadratic approximation) and, aside from a prefactor,
yields
\begin{equation}
dw\sim\exp \frac{1}{T}\int d^3x\left
[+\sqrt{m_+^2S^2(x)+m_-^2P^2(x)n_3^2}+m_+S(x)\right ]dn_3\, .
\label{corrdist}
\end{equation}
Although we could analyze this result it is simpler to consider the
situation where $|m_-P/m_+S|<1$. Keeping only the first term in the
expansion of the square root we obtain (ignoring, in the case $S(x)$
is positive, terms not depending on $n_3$)
\begin{equation}
dw\sim\exp\left [\frac{1}{2T}\frac{m_-^2}{m_+}\int
d^3x\frac{P^2(x)}{|S(x)|}n_3^2\right ]dn_3\, .\label{approxdist}
\end{equation}
Remembering that $f=n_3^2$ we find
\begin{equation}
dw=N(A)\, e^{Af}\frac{df}{2\sqrt{f}}\, ,\label{finaldist}
\end{equation}
where
\begin{equation}
A=\frac{1}{2T}\frac{m_-^2}{m_+}\int d^3x\frac{P^2(x)}{|S(x)|}\, ,
\label{defA}
\end{equation}
and the normalization factor
\begin{equation}
N^{-1}(A)=\int_0^1 dx e^{Ax^2}\, .\label{normalization}
\end{equation}
Evidently the change in the distribution is important only if $A$ is
large enough. In this case the distribution (\ref{finaldist})
has a minimum at $f=1/2A$ and, contrary to the situation described by
(\ref{fdist}), grows as $f$ approaches 1. For $A>>1$ an
approximate
evaluation of (\ref{normalization}) yields
\begin{equation}
dw\simeq Ae^{-(1-f)A}\frac{df}{\sqrt{f}}\, .
\end{equation}
This distribution has a peak at $f=1$ and is enhanced near that value
by a factor $2A$ over that of (\ref{fdist}) making anti-Centauros more
probable.

We shall now try to estimate possible values for $A$. While $|S(x)|$
presumably coincides with the quark density $\rho(x)$, $P(x)$ can be
represented in the form
\begin{equation}
P(x)=\xi R \mbox{\boldmath $\sigma$}(x)\cdot\mbox{\boldmath $\nabla$}
\rho(x)\, .\label{Pdef}
\end{equation}
Here \mbox{\boldmath $\sigma$}(x) is some spin density, R is a
characteristic linear size of the effective volume (or characteristic
time before hadronization) and $\xi$ is a constant, probably smaller than
one.

Integrating (\ref{defA}) we get
\begin{equation}
\int d^3x\frac{P^2(x)}{|S(x)|}=
4\pi R^2r\frac{\xi^2\rho^2R^2}{r^2}\frac{1}{\rho}=
\frac{4}{3}\pi R^3\frac{3R}{r}\xi^2\rho\, .
\end{equation}
We use $r$ as a characteristic length for the gradient; this variation
in density is likely to be
confined to the surface of the quark matter produced in the collision.
We assume that the volume over which $P(x)$ does not vanish is $4\pi
R^2r$. The spin densisities are averaged approximately to unity. Thus
for the parameter $A$ we have:
\begin{equation}
A=\frac{\xi^2}{2T}\
\frac{m_-^2}{m_+}\frac{3R}{r}N\simeq\frac{1}{70}\frac{R}{r}\xi^2N\, .
\end{equation}
Here $N=4\pi R^3\rho/3$ is the number of quarks produced. We believe
one could expect $N\geq 200$ in a sphere of $R\simeq 3$ fm (note that
for the vacuum $\rho=<{\bar q}q>=2\ \mbox{\rm fm}^{-3}$, so that
$N\simeq 200$). For $R/r\simeq 5$ we find $A\sim 15\xi^2$ and for
$\xi\geq 0.25$ $A$ will be sizable enough to enhance the probability
of anti-Centauros. Note that $\xi\leq 0.8$ is required for the
approximation in going from (\ref{corrdist}) to (\ref{approxdist}). We
are well aware of the crudeness of these estimates and the purpose of
this exercise was only to show that values of $A\geq 1$ are not excluded.

The whole change in the distribution of neutrals is due to the
violation of isotopic spin invariance; the parameter $A$ in
(\ref{finaldist}) is proportional to $(m_u-m_d)^2$. Can we claim that
the anti-Centauro events are caused by the mass difference of light
quarks?\\
{}

We would like to thank J.\ D.\ Bjorken for interesting discussions.
One of us (A.\ A.) would like to thank the Physics Department of the
University of California at Irvine for warm hospitality. This
research was supported in part by the National Science Foundation
under Grant PHY-9208386.

\end{document}